\documentclass[twocolumn,showpacs,preprintnumbers,amsmath,amssymb,superscriptaddress]{revtex4}
\usepackage{graphicx}% Include figure files
\usepackage{dcolumn}% Align table columns on decimal point
\usepackage{bm}% bold math

\begin{document}

\title{A double-dot quantum ratchet driven by an independently biased
quantum point contact}
\author{V.S.~Khrapai}
\affiliation{Center for NanoScience and Department f$\ddot{\text{u}}$r Physik,
Ludwig-Maximilians-Universit$\ddot{\text{a}}$t, Geschwister-Scholl-Platz 1,
D-80539 M$\ddot{\text{u}}$nchen, Germany} \affiliation{Institute of Solid State
Physics RAS, Chernogolovka, 142432, Russian Federation}
\author{S.~Ludwig}
\affiliation{Center for NanoScience and Department f$\ddot{\text{u}}$r Physik,
Ludwig-Maximilians-Universit$\ddot{\text{a}}$t, Geschwister-Scholl-Platz 1,
D-80539 M$\ddot{\text{u}}$nchen, Germany}
\author{J.P.~Kotthaus}
\affiliation{Center for NanoScience and Department f$\ddot{\text{u}}$r Physik,
Ludwig-Maximilians-Universit$\ddot{\text{a}}$t, Geschwister-Scholl-Platz 1,
D-80539 M$\ddot{\text{u}}$nchen, Germany}
\author{H.P.~Tranitz}
\affiliation{Institut f$\ddot{\text{u}}$r Experimentelle und
Angewandte Physik, Universit$\ddot{\text{a}}$t Regensburg, D-93040
Regensburg, Germany}
\author{W.~Wegscheider}
\affiliation{Institut f$\ddot{\text{u}}$r Experimentelle und Angewandte Physik,
Universit$\ddot{\text{a}}$t Regensburg, D-93040 Regensburg, Germany}

\begin{abstract} We study a double quantum dot (DQD) coupled to a strongly
biased quantum point contact (QPC), each embedded in independent
electric circuits. For weak interdot tunnelling we observe a
finite current flowing through the Coulomb blockaded DQD in
response to a strong bias on the QPC. The direction of the current
through the DQD is determined by the relative detuning of the
energy levels of the two quantum dots. The results are interpreted
in terms of a quantum ratchet phenomenon in a DQD energized by a
nearby QPC.
\end{abstract}
\pacs{73.63.Kv, 73.23.-b} \maketitle

In the absence of spatial symmetry, directed particle flow is
possible in a system subjected to fluctuations, without any
externally applied bias. The second law of thermodynamics requires
the fluctuations to be nonequilibrium~\cite{feynman}, i.e. a
directed flow only occurs in response to external energy supply.
Systems possessing current because of broken spatial symmetry, so
called ratchets, appear in a variety of examples from biological
systems to SQUIDs~\cite{reimann}.

In mesoscopic semiconductors, ratchet-type systems have been realized on the
basis of a two-dimensional electron gas in GaAs/AlGaAs
heterostructures~\cite{linkescience}. Spatial asymmetry results in rectified
current in periodic ratchets~\cite{linkescience}, single quantum
dots~\cite{linkeprb} and ballistic rectifiers~\cite{song}. Competition of
classical motion and quantum tunnelling can cause a crossover from a classical
to a quantum ratchet as the temperature is
decreased~\cite{Haenggi,linkescience}.

A double quantum dot (DQD) with internally broken symmetry, in
respect to its charge distribution, can operate as a quantum
ratchet. For weak interdot tunnelling, detuning of the quantum
dots' energy levels results in localization of an electron in one
dot, so that the elastic tunnelling to the other dot is
energetically forbidden. Externally supplied energy quanta can
promote inelastic interdot tunnelling, leading to a net current
flow through the DQD, as observed, e.g., in photon-assisted
tunnelling (PAT) experiments~\cite{vanderwiel}.

In this work a novel dynamic interaction effect between a DQD and an
independently biased quantum point contact (QPC) is reported. We observe a
finite current through the DQD in response to a strong bias on the QPC. The
results are interpreted in terms of a quantum ratchet phenomenon.

\begin{figure}[t]\vspace{2mm}
\scalebox{0.4}{\includegraphics[clip]{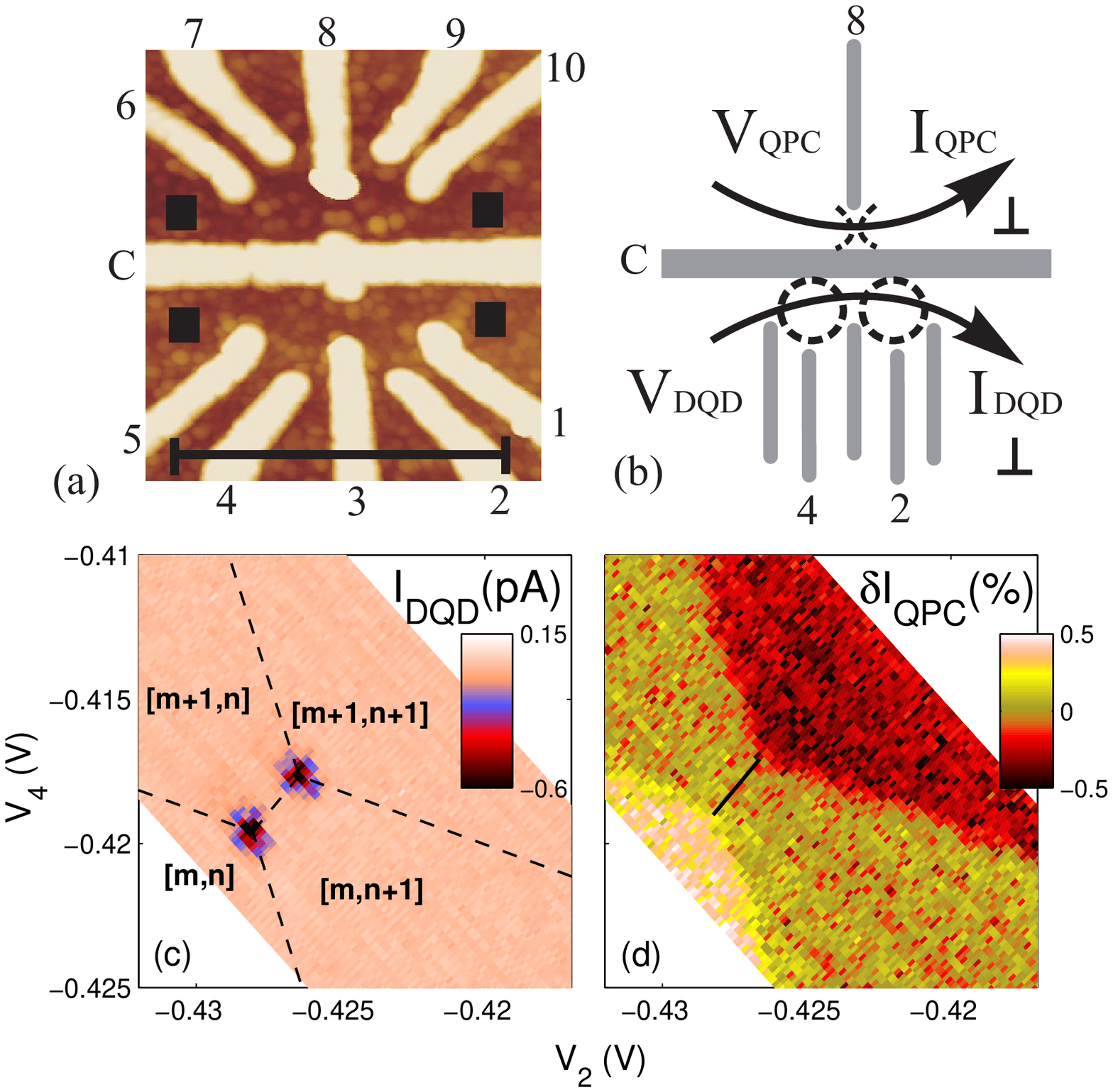}}\caption{(a) AFM
micrograph of the sample. Metal surface gates have a light color.
Black squares mark source and drain regions. The black scale bar
marks a length of $1\,\mu\text{m}$. (b) schematic layout of the
device as used for all presented measurements. (c) DQD stability
diagram for $V_\text{DQD}=-20\,\mu\,\text{V}$ and an unbiased QPC.
The axes show gate voltages on the plunger gates 2 and 4. (d) Same
area of the stability diagram as in (c) detected with the QPC for
$V_\text{QPC}=-1.45$~mV. A linear trend is subtracted from the
$I_\text{QPC}$ data. In (c) and (d) the QPC gate voltages are
$V_8=-0.597\,\text{V}$ and $V_7=V_9=-0.55\,\text{V}$.
$V_6=V_{10}=0$ throughout the paper. The black lines in (c) and
(d) are guides to the eyes (see text).
\label{fig1}}\vspace{-0.1in}
\end{figure}

Our samples are prepared on a GaAs/AlGaAs heterostructure
containing a two-dimensional electron gas $90$~nm below the
surface, with electron density $n_S=2.8\times10^{11}$~cm$^{-2}$
and mobility $\mu=1.4\times10^6$~cm$^2/$Vs at a temperature of
1.5~K. The AFM micrograph of the split-gate nanostructure,
produced with e-beam lithography, is shown in Fig.~\ref{fig1}a.
The negatively biased central gate C is used to divide the
electron system into two parts, so that direct leakage current
between them is absent. The DQD is defined and controlled by
negative voltages applied to the gates 1-5. Both serially coupled
quantum dots possess single-particle level spacings on the order
of $100~\mu$eV and charging energies of about 1.5~meV. The DQD has
small tunnel couplings of $\Gamma\approx40~\mu$eV to the leads and
a much smaller interdot tunnel coupling of $t_0\sim0.1~\mu$eV. The
QPC is formed on the opposite side of gate C by use of gate 8 as
well as side gates 7 and 9. The one-dimensional (1D) subband
spacing of the QPC is $\Delta E_S\approx4$~meV and the energy
window for opening a 1D subband is approximately 1~meV
wide~\cite{glazmann}.

A schematic layout of the device is shown in fig.~\ref{fig1}b.
Separated circuiting allows both, independent biasing and a
simultaneous current measurement of the DQD and the QPC. In each
circuit the source bias is applied to the left lead while the
right leads are grounded, so that positive (negative) current
corresponds to electrons moving from the right to the left (or
vice versa). We checked that the results are independent of the
particular grounding configuration. The experiments are performed
in a dilution refrigerator at an electron temperature below
150~mK. Dc current is measured using low noise current-voltage
amplifiers. Differential conductance data are obtained by
numerical derivation of the dc signal.

First, we discuss a conductance measurement of the DQD and
demonstrate the operation of the QPC as a charge
sensor~\cite{field}. Fig.~\ref{fig1}c shows a color scale plot of
current $I_\text{DQD}$ flowing through the DQD in dependence of
voltages on gates 2 and 4. Here, a small source-drain bias of
$V_\text{DQD}=-20~\mu$V is applied to the DQD, while the QPC is
kept unbiased. Gate voltages $V_2$ and $V_4$ predominantly control
the occupation of electronic states in the right and left dots,
respectively, allowing to scan a two-dimensional stability diagram
of the DQD~\cite{vanderwiel}. As expected for weak interdot
tunnelling, $I_\text{DQD}$ is non-zero only in the vicinity of the
so-called triple points, where Coulomb blockade is lifted and
electron-like or hole-like resonant tunnelling
occurs~\cite{vanderwiel}. Dashed guide lines in fig.~\ref{fig1}c
mark the boundaries between the regions of different ground state
charge configurations of the DQD. The charge configuration of the
DQD is denoted as [m,n], corresponding to m (n) electrons
occupying the left (right) dot.

In fig.~\ref{fig1}d we plot the current $I_\text{QPC}$ flowing
through the biased QPC for the same part of the DQD stability
diagram as shown in (c). $I_\text{QPC}$ increases stepwise each
time one electron leaves the DQD, as a result of electrostatic
interaction between the electrons traversing the QPC and those
localized in the DQD~\cite{field}. Because of the symmetric device
geometry (fig.~\ref{fig1}a), no change of $I_\text{QPC}$ is
observed when repositioning of an electron between the two quantum
dots occurs (across the solid line in
fig.~\ref{fig1}d)~\cite{marcus}.

\begin{figure}[t]\vspace{2mm}
\scalebox{0.7}{\includegraphics[clip]{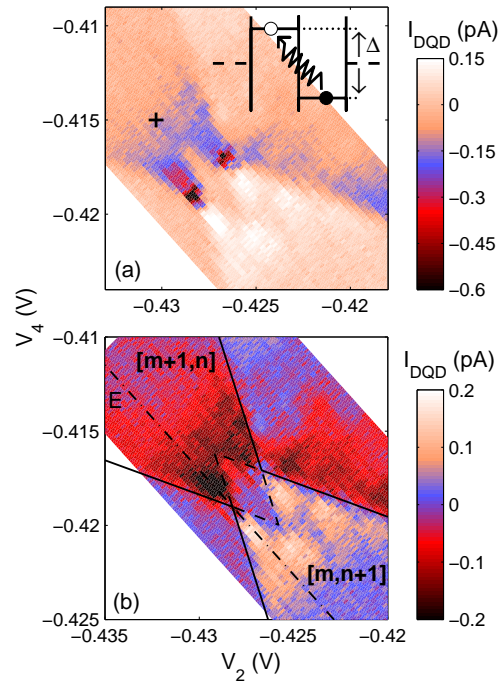}} \caption{Current
through the DQD while the QPC is biased with
$V_\text{QPC}=-1.45$~mV. The stability diagram area and gate
voltages are the same as in fig.~\ref{fig1}. (a) -- Raw data
measured at $V_\text{DQD}=-20\mu$V. (b) -- The same data after
subtraction of the resonant tunnelling contribution at the triple
points (see text). The solid lines in (b) mark boundaries between
regions of stable charge configurations. Dashed and dashed-dotted
lines are explained in the main text. Inset: a sketch of an
inelastic interdot tunnelling process involving absorption of an
energy quantum. Horizontal dashed lines mark the leads chemical
potential. The ground and excited electron states are marked by a
filled versus an empty circle,
respectively.\label{fig2}}\vspace{-0.1in}
\end{figure}

In the following we report on a new dynamic effect of a QPC-driven
current through a DQD. We bias the QPC at $V_\text{QPC}=-$1.45~mV
while leaving the DQD at the small bias of
$V_\text{DQD}=-20~\mu$V~\cite{remark1}. The dc conductance of the
QPC is adjusted to approximately $0.5\,G_0$, where $G_0=2$e$^2$/h
is the conductance quantum. Fig.~\ref{fig2}a
shows the raw data of $I_\text{DQD}$ for the same part of the
stability diagram as characterized before (fig.~\ref{fig1}c).
Remarkably, away from the triple points, in the regime of the
ground state Coulomb blockade, we observe a finite current through
the DQD, driven by the QPC bias
(compare~figs.~\ref{fig1}c~and~\ref{fig2}a). Fig.~\ref{fig2}b
shows $I_\text{DQD}$ for the effectively unbiased
DQD~\cite{remark1}, obtained after subtraction of the contribution
of resonant tunnelling at the triple points from the raw data
(fig.~\ref{fig2}a). The DQD signal changes abruptly at the
boundaries of the stability diagram (solid lines in
fig.~\ref{fig2}b) and at the edges of a diamond-shaped region
between the triple points (surrounded by dashed lines continuing
the solid lines). Within this diamond $I_\text{DQD}$ is close to
zero. Otherwise $I_\text{DQD}$ changes smoothly within each area
of a fixed ground state configuration. The direction of
$I_\text{DQD}$ is determined by the DQD charge configuration.
Note, that the current driven through the DQD is much smaller than
the QPC driving current (e.g. $I_\text{DQD}\sim0.5\,\text{pA} \ll
I_\text{QPC}\sim50\,\text{nA}$ for the data presented in
Fig.~\ref{fig2}). These features are not specific to the triple
points shown here, but periodically repeated throughout a broad
region of the stability diagram.

The observed dependence of the current driven through the DQD on
its ground state configuration can be understood in terms of
inelastic tunnelling, similar to PAT~\cite{vanderwiel}. For the
case of the ground state configuration [m,n+1], this process is
schematically shown in the inset to fig.~\ref{fig2}a. Initially
localized in the right dot, the highest energy electron can
resonantly absorb an energy quantum and tunnel to the left dot
(transition [m,n+1] $\rightarrow$ [m+1,n] in the DQD
configuration). The energy difference between the two charge
configurations is called the DQD asymmetry energy ${\Delta\equiv
E_{m+1,n}-E_{m,n+1}}$. The resonance condition requires that the
energy quantum is equal to the absolute value of the asymmetry
energy $|\Delta|$. In our device, the DQD relaxes back to the
ground state mainly via the dot-lead tunnelling of electrons,
because $\Gamma\gg t_0$. The excited electron escapes to the left
lead, whereas another electron enters the right dot from the right
lead, resulting in the observed net current through the DQD. Note,
that compared to the process described above, more probable are
inelastic processes involving ionization of one quantum dot
towards its adjacent lead~\cite{onac}, followed by recharging from
the same lead. These charge fluctuations, however, do not result
in a net current.

The relevance of the described current transfer mechanism is
confirmed by another observation. Within the diamond shaped region
bounded by dashed lines in fig.~\ref{fig2}b, both involved DQD
configurations, [m,n+1] and [m+1,n], are Coulomb
blockaded~\cite{vanderwiel}. This results in a suppression of the
inelastic DQD current, as observed experimentally
(fig.~\ref{fig2}).

\begin{figure}[t]\vspace{2mm}
\scalebox{0.45}{\includegraphics[clip]{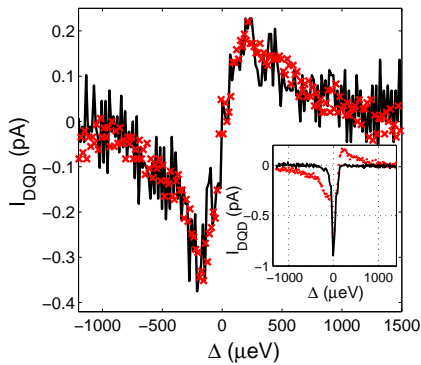}}
\caption{Inelastic current $I_\text{DQD}$ for
$V_\text{QPC}=-1.55\,\text{mV}$ (symbols) and $3.3 \times
I_\text{DQD}$ for $V_\text{QPC}=-1.15\,\text{mV}$ (solid line),
after substraction of the elastic contribution
$I_\text{DQD}(V_\text{QPC}=0)$, as a function of $\Delta$. The
data are measured along line E in fig.~\ref{fig2}b. Gate voltages
defining the QPC are the same as for fig.~\ref{fig1}. Inset: Raw
data measured at ${V_\text{QPC}=0}$ (line) and -1.45~mV (symbols),
and ${V_\text{DQD}=-20~\mu}$V. \label{fig3}}\vspace{-0.1in}
\end{figure}

Inelastic tunnelling processes in a DQD can be used to probe an excitation
spectrum~\cite{aguado,fujisawa}. In fig.~\ref{fig3}
the inelastic contribution to the DQD current $I_\text{DQD}$ is
plotted for two negative values of $V_\text{QPC}$ as a function of
$\Delta$ (along line E in fig.\ref{fig2}b, where gate voltages
have been converted to asymmetry energy~\cite{vanderwiel}). For
clarity, the resonant tunnelling peak at $\Delta=0$, measured at
$V_\text{QPC}=0$, has been subtracted from the data (see the raw
data in the inset of fig.~\ref{fig3}). The lower bias data (solid
line) are multiplied by a factor of 3.3. In this way, the two
curves scale onto a single one, suggesting identical $\Delta$
dependencies. $I_\text{DQD}(\Delta)$ is a nearly antisymmetric
function. The transition between negative and positive current at
$\Delta=0$ has a width of about $150\,\mu\text{eV}$, much broader
than the level width $\Gamma$. Hence, the shape of
$I_\text{DQD}(\Delta)$ at small $\Delta$ might be related to the
low energy fall-off in the excitation spectrum. For
$|\Delta|\gtrsim 1\,\text{meV}$ $I_\text{DQD}$ vanishes,
suggesting a 250~GHz bandwidth of the excitation spectrum. Note,
that the charging energy, which limits the bandwidth of a
DQD-based detector, is considerably larger. The broadband spectrum
of the DQD excitation results in a sharp contrast between the data
in figs.~\ref{fig2}~and~\ref{fig3} and conventional PAT
experiments~\cite{vanderwiel,marcus}.

A finite QPC-driven current through the DQD is only detected when
the highest energy electron is localized in one of the dots
($\Delta\ne0$). Thus, an internal asymmetry of the DQD is needed
to cause a dc current resulting from single electron tunnelling
processes between quantized energy levels. This demonstrates that
the DQD is a quantum ratchet. The QPC acts as non-equilibrium
energy source driving the DQD quantum ratchet.

\begin{figure}\vspace{2mm}
\scalebox{0.52}{\includegraphics[clip]{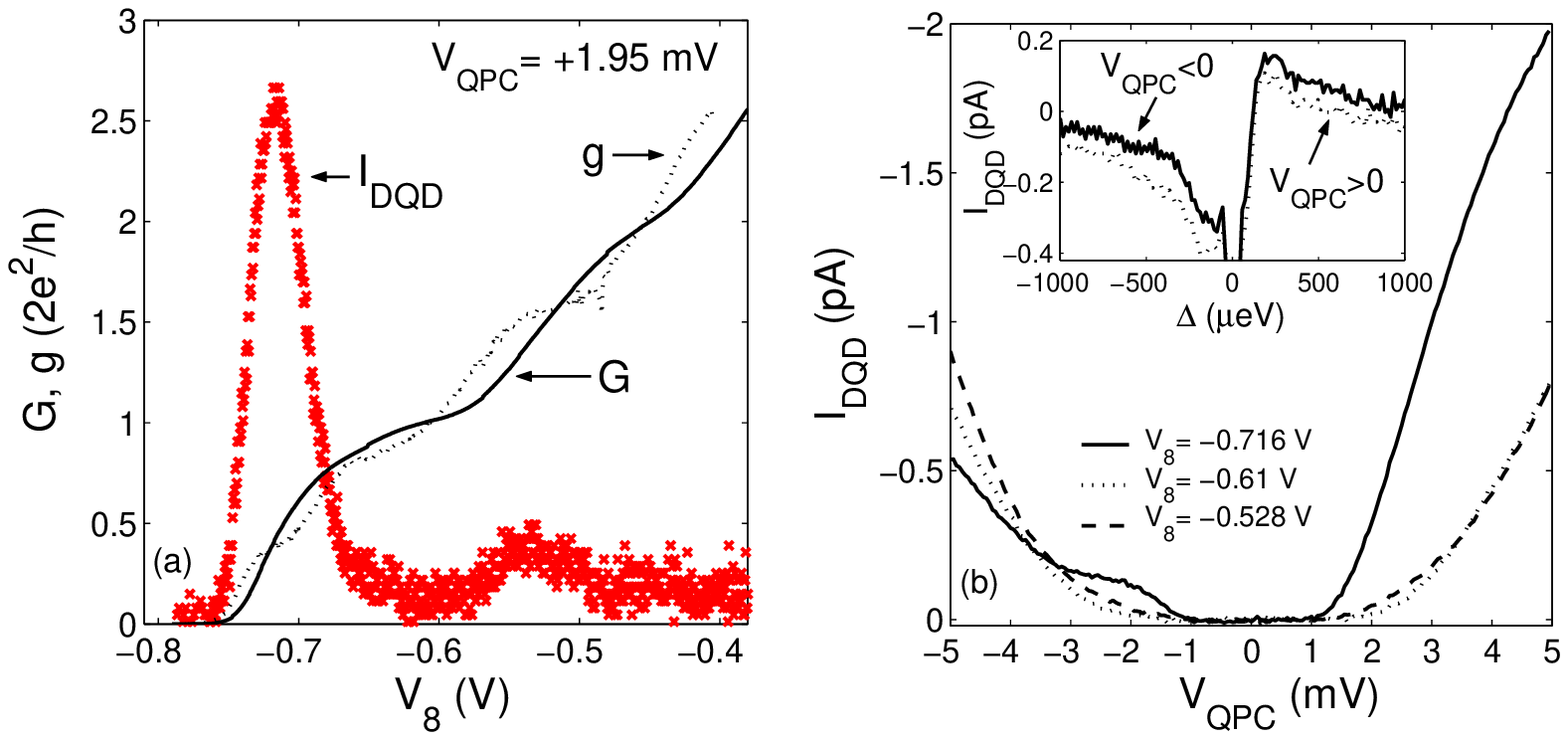}} \caption{((a)
$I_\text{DQD}$ (a.u.), dc conductance ($G$) and differential
conductance ($g$) of the QPC for $V_\text{QPC}=1.95\,\text{mV}$ as
a function of the QPC gate voltage $V_8$ ($V_7=V_9=0$). (b)
$I_\text{DQD}$ as a function of $V_\text{QPC}$ for 3 values of
$V_8$, corresponding to the two local maxima of $I_\text{DQD}$ in
(a) and the minimum in between. In both (a) and (b)
${\Delta=-450~\mu \text{eV}}$. The according position in the
stability diagram is marked in fig.~\ref{fig2}a by a black cross.
Inset: $I_\text{DQD}$ measured along line E in fig.~\ref{fig2}b
for $V_\text{QPC}=\pm1.45\,\text{mV}$. All current data shown in
fig.~\ref{fig4} are raw data.\label{fig4}}\vspace{-0.1in}
\end{figure}

In the following we study the ratchet driving mechanism by
changing the transmission and source bias of the QPC, while the
DQD ratchet is adjusted to a fixed asymmetry energy
$\Delta=-450~\mu$eV (as marked by a black cross in
fig.~\ref{fig2}a). Figure~\ref{fig4}a
shows the measured current $I_\text{DQD}$ as a function of the
gate voltage $V_8$, controlling the transmission of the QPC at a
fixed source bias of $V_\text{QPC}=1.95\,\text{mV}$. Also shown
are the conductance ($G=I/V$) and the differential conductance
($g=\text{d}I/\text{d}V)$ of the QPC. The plateaus at
$G=G_0$~and~$G=2G_0$ correspond to the onset of full transmission
in the first two 1D subbands~\cite{glazmann}. The regions of
partial transmission between the plateaus are broadened because of
the applied bias $V_\text{QPC}$. Within these regions the
differential conductance shows so-called half-plateaus at
$g\approx0.4\,G_0$~and~$g\approx1.6\,G_0$~\cite{cronenwett,glazmann}.
The ratchet current $I_\text{DQD}$ has two local maxima at
voltages corresponding to the half-plateaus in $g$. The first
maximum above the onset of the QPC conduction is much more
pronounced than the second one. $I_\text{DQD}$ is minimal in case
of fully transmitting 1D subbands of the QPC. This non-monotonous
dependence of $I_\text{DQD}$ on the QPC transmission excludes the
energy dissipation of electrons in the leads of the QPC as the
ratchet driving mechanism. We conclude, that the energy source is
determined by local dissipation in the QPC.

Figure~\ref{fig4}b shows $I_\text{DQD}$ as a function of
$V_\text{QPC}$ for QPC gate voltages tuned to the positions of the
two current maxima in fig.~\ref{fig4}a or the minimum in between.
As expected for a ratchet, $I_\text{DQD}$ is negative, regardless
of the sign of $V_\text{QPC}$. Strikingly, at small bias
${|\text{V}_\text{QPC}|\lesssim1\,\text{mV}}$ the current through
the DQD vanishes. This onset value is independent of $\Delta$
(refer to the scaling behavior observed in fig.~\ref{fig3}). As
$V_\text{QPC}$ increases the two maxima broaden, and for
$|V_\text{QPC}|\gtrsim\Delta E_S$ the minimum in between
disappears, explaining the dotted curve in fig.~\ref{fig4}b.

We now discuss a possible ratchet driving mechanism, resulting
from the strongly biased QPC. The condition of partial
transmission in 1D subbands of the QPC, necessary for
$I_\text{DQD}\ne0$ (fig.~\ref{fig4}a), signals the importance of
shot noise~\cite{blanter}. Voltage fluctuations originating from
current shot noise is a known excitation mechanism in on-chip
detection schemes~\cite{aguado,onac}. However, for the DQD
ratchet, in this case the onset QPC bias is expected to depend on
the DQD asymmetry energy as $V_\text{QPC}=\Delta/e$, in contrast
to our experimental results~\cite{remark2}. We believe that the
energy relaxation of electrons traversing the QPC is involved.
Occupation number fluctuations, caused by shot noise, lift the
Fermi degeneracy of the current carrying 1D electron states. At
high bias, unoccupied current carrying states exist within an
energy window of about 1~meV near the 1D subband bottom in our
QPC. This is expected to enhance the energy relaxation rate of
electrons within the QPC. Qualitatively, this could account for
the observed nonmonotonic dependence of the ratchet current upon
the QPC transmission (fig.~\ref{fig4}a). However, quantitatively,
the relative height of the two current maxima (fig.~\ref{fig4}a),
as well as the suppression of $I_\text{DQD}$ at
${|\text{V}_\text{QPC}|\lesssim1\,\text{mV}}$ (fig.~\ref{fig4}b),
is lacking explanation. The energy quanta, emitted by the QPC
electrons and absorbed by the electrons in the DQD, could be short
wave length acoustic phonons, long wave length photons or
1D-plasmons. A further discrimination is difficult, because close
to the 1D subband bottom no strict constraints are imposed by the
momentum conservation law~\cite{matveev}.

Finally, we mention an additional contribution to the DQD current,
which is not related to the ratchet phenomenon. The inset of
fig.~\ref{fig4}b plots the measured $I_\text{DQD}$ as a function
of $\Delta$ for positive versus negative
$V_\text{QPC}=\pm1.45$~mV. Clearly, apart from the ratchet current
contribution (antisymmetric in $\Delta$), an additional negative
and $\Delta$-independent contribution to $I_\text{DQD}$ is seen
for $V_\text{QPC}>0$. This contribution to $I_\text{DQD}$, which
is always opposite to the current in the QPC, is responsible for
the bias asymmetry of one curve seen in fig.~\ref{fig4}b (solid
line). It can be attributed to a drag-type effect, operating
independently of the ratchet phenomenon near the QPC pinch-off, as
e.g. phonon-mediated adiabatic pumping~\cite{levinson} or
statistically asymmetric voltage noise~\cite{luczka}.

In summary, we find a current through a DQD, driven by a strongly
biased nearby QPC. We demonstrate that the DQD acts as a quantum
ratchet, energized by the current through the independently biased
QPC. The experimental results are qualitatively consistent with
inelastic relaxation of electrons in partly transmitting 1D
channels of the QPC.

The authors are grateful to V.T.~Dolgopolov, A.W.~Holleitner,
C.~Strunk and F.~Wilhelm for valuable discussions and to
D.~Schr$\ddot{\text{o}}$er and M.~Kroner for technical help. We
thank the DFG via SFB 631 and VSK the Alexander~von~Humboldt
foundation for support.


\begin{thebibliography}{apssamp}
\bibitem{remark1} Direct measurements at $V_{\rm DQD}=0$ are superimposed by a signal caused
by rectification of 50~Hz noise ($\simeq 10\mu$V~rms). This we
avoid by using a finite bias of $V_{\rm DQD}=-20\,\mu$V. We find
the QPC-driven curent through the DQD to be independent of bias
for $|V_{\rm DQD}|<50\,\mu$V.
\bibitem{remark2} For the parameters of our system, we
estimate~\cite{aguado} that the current $I_\text{DQD}$, caused by
the shot noise voltage fluctuations on the QPC, is at least 3
orders of magnitude smaller than the experimentally found ratchet
current.
\end{thebibliography}
\end{document}